\documentclass[pdflatex,sn-basic]{sn-jnl}

\setcitestyle{aysep={,},notesep={,}}

\usepackage{graphicx}%
\usepackage{multirow}%
\usepackage{amsmath,amssymb,amsfonts}%
\usepackage{amsthm}%
\usepackage{mathrsfs}%
\usepackage[title]{appendix}%
\usepackage{xcolor}%
\usepackage{textcomp}%
\usepackage{manyfoot}%
\usepackage{booktabs}%
\usepackage{algorithm}%
\usepackage{algorithmicx}%
\usepackage{algpseudocode}%
\usepackage{listings}%

\theoremstyle{thmstyleone}%
\theoremstyle{thmstyletwo}%

\theoremstyle{thmstylethree}%

\newcommand{\YS}[1]{{\color{black}  #1}}

\begin{document}

\title[Article Title]{From Accessibility to Allocation: An Integrated Workflow for Land-Use Assignment and FAR Estimation}

\author*[1]{\fnm{Yue} \sur{Sun}}\email{ys954@cornell.edu}

\author[2]{\fnm{Ryan} \sur{Weightman}}

\author[1]{\fnm{Yang} \sur{Yang}}
\author[1]{\fnm{Anye} \sur{Shi}}
\author[3]{\fnm{Timur} \sur{Dogan}}
\author[4]{\fnm{Samitha} \sur{Samaranayake}}

\affil*[1]{\orgdiv{System Engineering Department}, \orgname{Cornell University}, \orgaddress{\city{Ithaca}, \postcode{14850}, \state{N.Y.}, \country{U.S.A.}}}

\affil[2]{\orgdiv{Center for Computational and Integrative Biology}, \orgname{Rutgers University - Camden}, \orgaddress{\street{Street}, \city{Camden}, \postcode{08102}, \state{N.J.}, \country{U.S.A.}}}

\affil[3]{\orgdiv{Environmental Systems Lab}, \orgname{Cornell University}, \orgaddress{\city{Ithaca}, \postcode{14850}, \state{N.Y.}, \country{U.S.A.}}}

\affil[4]{\orgdiv{School of Civil Environmental Engineering}, \orgname{Cornell University}, \orgaddress{\city{Ithaca}, \postcode{14850}, \state{N.Y.}, \country{U.S.A.}}}


\abstract{Urban land use and building intensity are often planned without a direct, auditable link to network accessibility, limiting ex-ante policy evaluation. This study asks whether multi-radius street centralities can be elevated from diagnosis to design lever to allocate land use and floor area in a transparent, optimization-ready workflow. We introduce a three-stage pipeline that connects configuration to program and intensity. First, multi-radius accessibility is computed on the street network and translated to blocks to provide scale-legible measures of reach. Second, these measures structure nested service basins that guide a rule-based placement of land uses with explicit priorities and minimum parcel footprints, ensuring reproducibility. Third, within each use, floor-area ratio (FAR) is assigned by an accessibility-weighted linear model that satisfies global construction totals while anchoring the average FAR, thereby tilting height toward better-connected blocks without pathological extremes. The framework supports multi-objective policy search via sampling and Pareto screening. Applied to a real urban district, the workflow reproduces corridor-biased commercial siting and industrial belts while concentrating intensity on highly connected blocks. Policy sampling via multi-objective screening yields Pareto-efficient plans that reconcile accessibility gains with deviations from target land- and construction-share structures. The contribution is twofold: methodologically, it translates familiar space-syntax measures into cluster-aware, rule-governed land-use and FAR assignment with explicit guarantees (scale-legible radii, parcel minima, and an average-FAR anchor); practically, it offers planners a transparent instrument for counterfactual testing and negotiated trade-offs at neighborhood–district–city scales.}

\keywords{Accessibility, land-use allocation, FAR, multi-objective optimization, design space exploration}



\maketitle

\section{Introduction}\label{intro}
\subsection{Motivation and Contributions}
Urban planning workflows often decouple three tightly linked choices—network configuration, land-use siting, and development intensity (FAR/height)—into sequential, siloed decisions. This fragmentation weakens our grasp of feedback along the configuration, program, and intensity chain: how spatial structure conditions natural movement and frontages, how uses anchor where intensity is credible, and how policy levers (radii, targets, priorities) jointly affect travel demand (e.g., vehicle miles traveled, VMT) and jobs–housing balance \citep{hillier1996space,ewing2010travel,cervero1996mixed,kockelman1997travel}. In the space-syntax “movement economy,” configuration shapes aggregate movement, which in turn shapes land uses and building forms reciprocally \citep{hillier1996space,hillier2005network,porta2006network,crucitti2006centrality}. Yet many generative land-change models either abstract away frontage-sensitive siting or treat intensity as exogenous envelopes: cellular automata and related transition models excel at raster growth but struggle at parcel/block logic and interpretability \citep{white1993cellular,clarke1997sleuth,batty1997cellular,white2000modeling}; procedural pipelines (e.g., CGA grammars, CityEngine) compactly encode form but rarely integrate siting and height to empirically grounded accessibility signals \citep{parish2001procedural,muller2006procedural,kelly2021cityengine}. Classical urban economics already imply that intensity responds to accessibility via bid-rent \citep{alonso1964location,muth1969cities,fujita1989urban}, yet the mainstream zoning practice is to fix intensity by code rather than co-determining it with multi-scale access and policy.\\

\noindent Against this backdrop, our inquiry proceeds from broad theory to a specific testable problem. We ask whether elevating accessibility from backdrop to driver can improve access-weighted intensity while holding land-use and construction-share deviations in check. We further probe at which travel radii and priority orders these trade-offs tighten or relax in ways consistent with everyday service basins \citep{hansen1959accessibility,geurs2004accessibility}. We then examine, under explicit multi-objective optimization, which mismatches become binding—land-use shares or construction shares, and how this reconfigures the Pareto frontier \citep{deb2002nsga2,satopaa2011kneedle}. Our proposed framework emphasizes a single, reproducible experiment that keeps configuration, allocation, and intensity in one loop. Methodologically, we compute choice/integration at user-selected radii and map segment scores to blocks via frontage-sensitive aggregation, so accessibility becomes the mode of allocation rather than a backdrop. We then implement a cluster-aware, priority-guided allocation that operationalizes planning intent (targets, fairness/compactness, minimum parcel thresholds) in readable steps—an interpretable alternative to black-box transition rules. Finally, we model FAR/height as accessibility-weighted intensity: meeting a global construction target while tilting height towards reach, thereby aligning delivered intensity with network advantage instead of only with statutory maxima. We explore the policy space by sampling radii, shares, weights, and priorities and by screening non-dominated plans; this explicitly displays how accessibility benefit (or conflicts) co-moves with land-use fit and jobs–housing indicators, enabling principled “knee” selection rather than single-best claims \citep{Hewitt02122022, McKay1979LHS}.\\

\begin{figure}[!h]
\begin{center}
\includegraphics[angle=0,
scale=0.25,
trim={0cm 0cm 0cm 0cm},clip]{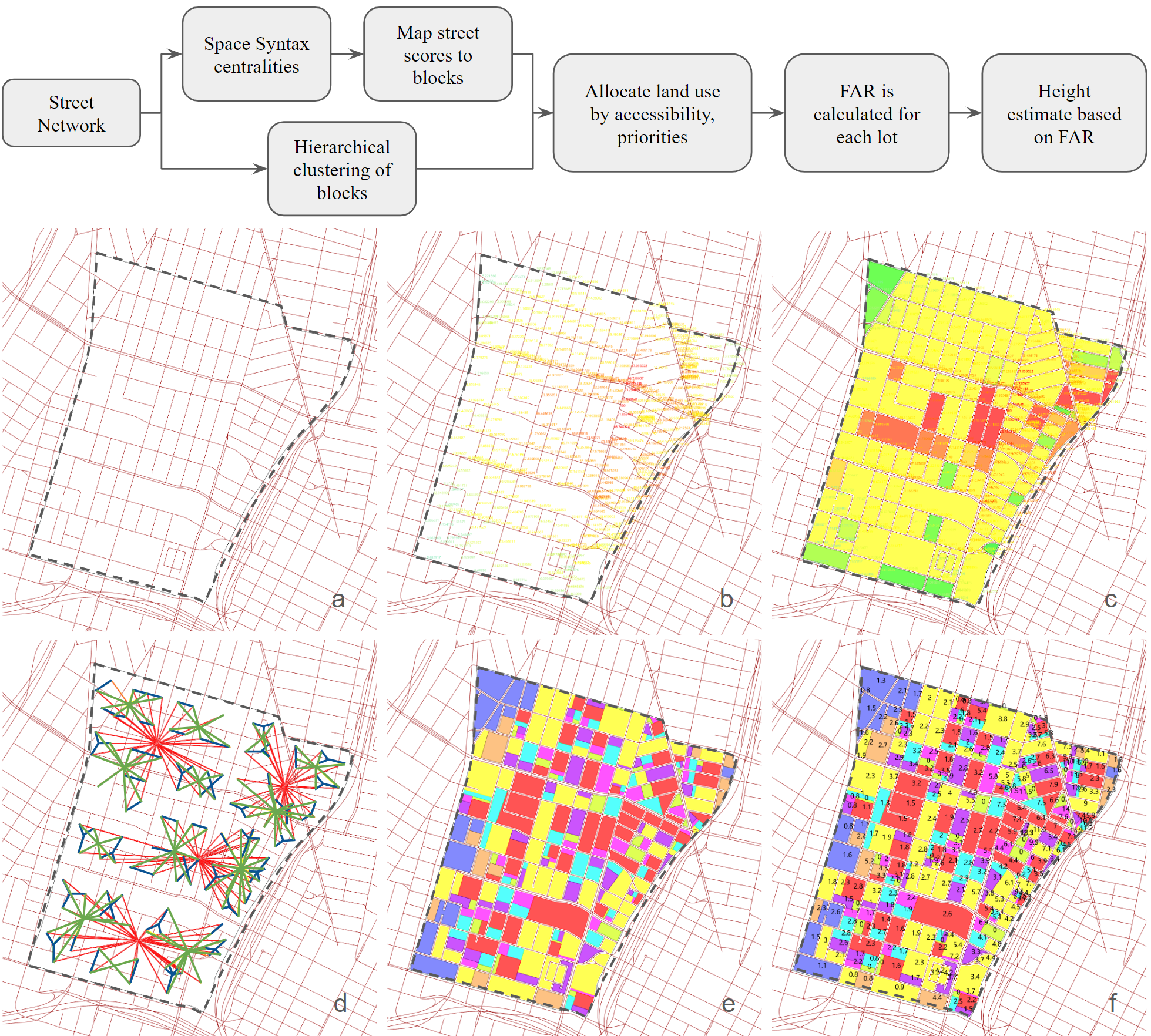}
\caption{The flow diagram and visualization of land use allocation and FAR calculation on site. a) The site with its original street network. b) Accessibility analysis for each tier of radius. c) Transferring segment accessibility scores to blocks. d) Hierarchical clustering of blocks for each tier. \YS{We are showing 3 tiers of clusters, therefore, 3 cluster colors in the plot.} e) Land use allocation. f) FAR calculation with each lot being assigned a FAR value.} 
\label{fig:flow_on_site}
\end{center}
\end{figure}

\noindent Taken together, we propose a framework capable of turning a set of often disconnected analyses into a usable planning instrument. It lets researchers probe the theorized chain—configuration $\leftrightarrow$ program $\leftrightarrow$ intensity through explicit counterfactuals, while giving practitioners an auditable way to test commitments at city, district, and neighborhood scales, precisely where guessing is costliest and transparent trade-offs matter most (Fig.~\ref{fig:flow_on_site}). Our contribution speaks to three ongoing conversations. First, in space-syntax and network-centric planning, \YS{we demonstrate how segment-level angular/metric scores can drive block-scale allocation while maintaining interpretability of each step (i.e., each rule is transparent and grounded in planning logic)}, and we do so across multiple, practitioner-legible radii. \YS{However, we acknowledge that the interaction of multiple simple rules across scales can yield complex, emergent outcomes not always predictable a priori. In other words, our framework improves interpretability relative to black-box models, but it does not eliminate complexity-it makes the sources of complexity explicit.} Second, in land-use allocation and urban growth modeling, we introduce a cluster-aware, rule-based allocator that retains much of the explanatory power of cellular automata and genetic algorithms approaches while remaining auditable—using priority queues, top-tier handling of "good" versus "bad" uses, and residual assignment to residential. Third, in intensity modeling, we bridge design practice and urban economics: linear accessibility-to-FAR mapping reallocates heights while preserving average intensity, capturing observed demand gradients alongside the realities of binding caps. By structuring the workflow as rules rather than opaque weights, the method invites disciplinary scrutiny (space-syntax theory; land-use-transport interaction traditions) and practice-side negotiation (service tiers, parcel minima, management of undesirable uses). \YS{In short, while we realize that each rule enhances auditability and their interactions may produce emergent outcomes not fully predictable without simulation, nonetheless, this positions accessibility as a design lever—one that makes trade-offs among land-use mix and intensity explicit, and can be tuned to local priorities within a transparent, reproducible allocation pipeline.}

\subsection{Background and Related Work}
Urban form has long been studied through computationally generative models designed to reproduce realistic city structures from a compact set of rules. These models typically consist of a workflow, a generative method, and a series of constraints \citep{shi2017review}. Approaches span shape grammars \citep{OzkarKotsopoulos2008IntroductionShapeGrammars,wang2020generative,duarte2011towards}, L-systems \cite{coelho2007expeditious,marvie2005fl,kelly2021cityengine}, agent-based simulations \citep{luca2007generative}, tensor-field tracing \citep{sun2023generative}, and cellular automata \citep{batty1997cellular,white1993cellular}. Together, these provide procedural or parametric pipelines (e.g., CityEngine, OSM-based workflows) for generating urban fabric that can serve as upstream input to accessibility analysis, allocation routines, and density estimation. Their shared strength is expressive power and rapid scenario synthesis. Our study inherits the procedural spirit (i.e., explicit rules, fast generation) but shifts emphasis to post-hoc auditability: each allocation step is encoded as an observable rule (tiering, priorities, minimum parcel sizes), so a planner can change inputs and see traceable effects.\\

\noindent A central concept linking these generative processes to urban performance is accessibility. Defined as the ease with which activities, opportunities, or services can be reached from a given location, accessibility captures the interaction between transport networks, land-use distributions, and the opportunities themselves. Unlike mobility, which emphasizes movement, accessibility foregrounds the connection between where opportunities are located and the effort or cost required to reach them \citep{sola2018understanding}. Historically, it has anchored urban economics and geography by showing how access to markets and services shapes land-use patterns \citep{norton1979relevance,getis1966christaller}. Modern measures span from simple proximity-based counts to impedance-weighted gravity formulations and location-based metrics that integrate land-use, transport, temporal, and individual components. In doing so, accessibility provides a unifying lens for analyzing equity, efficiency, and infrastructure impacts. Its virtue is direct interpretability and policy resonance and the limitation is dependence on exogenous impedance and activity data, which are often uncertain or unavailable at early design stages. Space syntax offers a configurational alternative which introduces configurational predictors of movement: angular integration and choice (segment-based closeness/betweenness under turning costs) predict pedestrian/circulation potentials from street geometry alone \citep{hillier1976space}. These measures parallel "centrality" in network science and have been empirically linked to pedestrian flows and co-location of land uses, highlighting how spatial configuration itself structures accessibility. Our pipeline computes both metric and angular centralities across user-specified radii and exposes them as analyzable vectors per segment, thereby unifying configurational and distance-based perspectives at multiple scales. The advantage is internal consistency; a limitation is that pure configurational proxies may miss demand heterogeneity, which we partially buffer via clustering and tiered priorities.\\

\noindent \YS{Modern accessibility measures span various scales, but typically analyses either fix a scale or treat each scale separately \citep{chaudhuri2013sleuth, huang2019delimiting, song2018improved}. Our approach explicitly combines multiple scales of street centrality in one workflow, allowing neighborhood-level and city-level accessibility to inform decisions concurrently. This multi-scale integration builds on space syntax traditions but extends them: whereas previous studies compute multi-radius metrics mostly for analysis or correlation, we use them as direct inputs to generative rules. This ensures that, for instance, both local-accessibility (e.g., 400m radius) and regional-accessibility (several km radius) can shape different aspects of the plan (land-use placement, intensity), a feature uncommon in earlier models.} Our allocation mechanism is deliberately legible. We (1) pass multi-radius segment scores to blocks, organizing block accessibility by “living-radius” tiers; (2) divide uses into “good” and “bad” queues to control early vs. late placement in the priority order; and (3) allocate within service clusters, where each cluster’s share is proportional to its remaining area. These choices answer practitioners’ recurring request for explanations they can debate and adjust, rather than opaque weights. The multi-scale specification of living radii address the need for emphasis on the scale difference. Scale and organization matter. Local radii better explain convenience retail/community services; district/city radii capture office/logistics comparisons. To avoid “winner-take-all” siting at top-access nodes, we group blocks into service clusters at each living-radius tier via hierarchical clustering; clusters then become the allocation unit. We enforce a simple dispersion guardrail—clusters are “available” only when the sum of remaining area in their blocks is at least above the specified thresholds—so allocations do not dissolve into slivers. This tiered logic resonates with proximity-planning arguments (e.g., 15-minute city \citep{smartcities4010006}), but with an explicit, testable mechanism: nested groups at multiple radii, top-down allocation from coarse to fine tiers, and parcel-size minima to maintain viability.\\

\noindent Where accessibility and network configuration establish a basis for evaluating potential, land-use allocation frameworks translate those potentials into explicit spatial patterns. Cellular automata models such as SLEUTH simulate urban growth using raster transition rules across inputs such as slope, land cover, exclusion, urbanization, transportation, and hillshade \citep{chaudhuri2013sleuth}. \YS{CA models offer fine-grained simulations but often act as "black boxes" with emergent behavior that is hard to trace to specific rules, and they may not incorporate street-network accessibility explicitly.} Demand-driven models like CLUE (Conversion of Land Use and its Effects) \citep{huang2019delimiting} emphasize competition among multiple land-use classes to capture regional planning dynamics. Optimization-based approaches such as MOLA (Multi-objective Land Allocation) extend this by allocating parcels to uses subject to multiple objectives and constraints \citep{song2018improved}. \YS{They can handle multiple objectives but typically require abstract encoding of planning goals and can be opaque in operation.} Our method aligns most closely with this optimization family, but implements a structured allocation routine using ordered land-use targets and block-level priority queues. \YS{Specifically, we propose an explicit rules-based allocator: we encode planning heuristics (e.g., prioritize certain uses on highly accessible sites, avoid fragmenting industrial lands) as transparent algorithms. Our framework retains interpretability/auditability – each step (tier allocation, clustering, etc.) is explainable – while still achieving complex outcomes. This aligns with procedural models (like CityEngine rule systems \citep{kelly2021cityengine}) in spirit, but uniquely, we tie the rules to accessibility metrics, marrying configurational analysis with allocation in a single system. Table \ref{table:ComparisonWithPriorWork} compares our work with key studies. To our knowledge, prior studies have not combined multi-scale network centrality with a rule-based allocation in this manner, especially with an eye toward optimization and policy exploration.}\\

\begin{sidewaystable}
\caption{Comparison of our multi-scale, explicit rules-based planning approach against existing studies and practices.}\label{table:ComparisonWithPriorWork}
\begin{tabular*}{\textheight}{@{\extracolsep\fill}ccccc}
\toprule%
Study & Approach & Key Features & Similarities to Our Work & Differences from Our Work \\
\midrule
\parbox{3cm}{SLEUTH \citep{chaudhuri2013sleuth, AsadiEtAl2022UrbanSprawlSimulationUrmia}}   & \parbox{2.5cm}{Cellular automata (CA) for urban growth simulation}   & \parbox{4.5cm}{Raster-based transitions using slope, land cover, exclusion, urbanization, transportation, and hillshade; self-modifying rules for diffusion, breed, spread, slope, and road gravity. Recent updates incorporate ANN for enhanced prediction.}  & \parbox{3.2cm}{Simulates spatial growth patterns; incorporates accessibility via transportation layers.} &  \parbox{4.5cm}{Single-scale, probabilistic transitions lack explicit rules or multi-radius nesting; treats intensity as exogenous; black-box self-modification reduces auditability vs. our rule-based priorities and parcel minima.}\\
\midrule
\parbox{3cm}{CLUE \citep{Balhas2025MultilevelParcelCA, huang2019delimiting}}  & \parbox{2.5cm}{Demand-driven CA for land-use competition} & \parbox{4.5cm}{Multi-scale allocation balancing supply-demand; integrates biophysical and socioeconomic drivers for transition probabilities. Recent extensions use multi-level neighborhoods for parcel-based effects.} & \parbox{3.2cm}{Multi-objective competition among uses; accessibility via location factors.} & \parbox{4.5cm}{Relies on statistical regressions for transitions, not explicit priorities; no endogenous FAR or space-syntax integration; less focus on block-scale interpretability vs. our tiered, cluster-aware rules.} \\
\midrule
\parbox{3cm}{MOLA \citep{HAJEHFOROOSHNIA2011254, GHARAIBEH20251179, Mutlu2020OptimizationMOLAGeneticAlgorithm}}  & \parbox{2.5cm}{Multi-objective optimization for land allocation} & \parbox{4.5cm}{Parcel-based assignment optimizing compactness, contiguity, and suitability under constraints; uses genetic algorithms or simulated annealing. Recent updates adopt GIS for urban growth modeling.} & \parbox{3.2cm}{Multi-objective Pareto search; explicit constraints for feasibility.} & \parbox{4.5cm}{Abstracts accessibility as exogenous suitability; no multi-scale basins or frontage-sensitive rules; optimization is black-box vs. our auditable priority queues and accessibility-weighted FAR.} \\
\midrule
\parbox{3cm}{CityEngine/CGA \citep{JIANG2024100795, muller2006procedural}}  & \parbox{2.5cm}{Procedural grammars for urban modeling} & \parbox{4.5cm}{Shape grammars (CGA) for rule-based generation of buildings and fabrics; integrates GIS data for parametric designs. Recent reviews emphasize AI integration for generative urban design.} & \parbox{3.2cm}{Procedural, rule-based generation; supports rapid scenario synthesis.} & \parbox{4.5cm}{Exogenous envelopes for intensity; limited accessibility integration (e.g., no multi-radius space syntax); focuses on form over allocation vs. our end-to-end chain from configuration to program/intensity.} \\
\midrule
\parbox{3cm}{Multiscale Accessibility and Urban Performance \citep{KaplanBurgOmer2022MultiscaleAccessibilityUrbanPerformance}}  & \parbox{2.5cm}{Space syntax with multi-scale network analysis} & \parbox{4.5cm}{Examines multiscale accessibility's relationship to urban performance; uses angular integration/choice across radii.} & \parbox{3.2cm}{Multi-scale configurational accessibility; links space to function.} & \parbox{4.5cm}{Diagnostic rather than generative; no explicit allocation rules or FAR optimization; lacks policy sampling vs. our design-lever use of radii/priorities for counterfactuals.} \\
\midrule
\parbox{3cm}{Generative Urban Design: A Systematic Review \citep{JIANG2024100795}}  & \parbox{2.5cm}{Review of generative methods with AI} & \parbox{4.5cm}{Surveys evolutionary optimization and deep models for urban fabric generation; focuses on efficiency in exploring solution spaces.} & \parbox{3.2cm}{Generative exploration of urban forms; multi-objective trade-offs.} & \parbox{4.5cm}{Emphasizes black-box AI (e.g., GANs) over explicit rules; single-scale or abstract accessibility; no nested basins or auditable heuristics vs. our transparent, rule-governed pipeline.} \\
\midrule
\parbox{3cm}{Multi-Objective Optimization for Smart Cities \citep{ChenChanSuDiao2025MOO_SmartCities_SLR}}  & \parbox{2.5cm}{Systematic review of MOO algorithms} & \parbox{4.5cm}{Bio-inspired methods (e.g., NSGA-II) for urban systems like transport and energy; Pareto frontiers for sustainability trade-offs.} & \parbox{3.2cm}{Multi-objective optimization; accessibility in some models as suitability.} & \parbox{4.5cm}{Aggregates methods without explicit rules; often single-scale; no integration of space syntax or endogenous FAR tilting vs. our accessibility-anchored, rule-governed pipeline.} \\
\midrule
\parbox{3cm}{A New Multi-Level Neighborhood Parcel-Based CA \citep{Balhas2025MultilevelParcelCA, CaiWang2020ThematicResolutionCAMarkov}}  & \parbox{2.5cm}{Enhanced CA for land-use allocation} & \parbox{4.5cm}{Multi-level neighborhoods for parcel-scale effects; simulates urban changes with improved neighborhood interactions.} & \parbox{3.2cm}{Multi-scale simulation; parcel-based allocation.} & \parbox{4.5cm}{Probabilistic CA transitions vs. explicit priorities; exogenous intensity; less auditability than our rule-based queues and minima for FAR assignment.} \\
\botrule
\end{tabular*}
\end{sidewaystable}

\noindent Intensity (FAR/height) allocation connects design practice to urban economics. A large economics literature shows that binding intensity controls and zoning frictions re-shape supply, prices, and spatial equilibria \citep{Supply, turner2007axial}. Yet in practice, FAR is often imposed exogenously rather than co-determined with access. Translating those insights into a tractable design tool, we specify a simple linear mapping from accessibility to FAR that preserves the user-specified average FAR while differentially rewarding more accessible blocks. It aims to keep the mechanism transparent (a one-line model a planner can reason about) while letting accessibility shift height envelopes toward likely demand foci. The FAR stage is implemented after land-use assignment and can be tuned by land-use-specific area targets and accessibility weights (Fig~\ref{fig:flow_on_site}).\\

\noindent Finally, policy objectives and evaluation criteria in the literature are likewise multi-objective. Beyond accessibility uplift, we use the proposed framework to explore jobs–housing balance to temper commuting growth and alignment with regulatory land-use share targets. Our evaluation follows that multi-objective tradition: we measure deviation from target land-use shares, construction-share consistency, and a jobs–housing penalty, then identify non-dominated policies and select a knee solution via utopia-point proximity. To sample policies efficiently we draw on well-established experimental-design and search heuristics—Latin hypercube sampling for broad, space-filling policy exploration and non-dominated sorting for Pareto extraction \citep{turner2007axial, HillierIida2005NetworkPsychUrbanMovement}.

\section{Methods}
\subsection{Workflow overview}
We propose a three-stage pipeline linking network accessibility to program and intensity. First, we build a multi-scale space-syntax graph and compute segment Choice/Integration (metric/angular), aggregate to blocks, and normalize per tier. Second, we cluster blocks by travel radii to form nested catchments, then allocate land uses top-down: coarse→fine tiers, "good" land uses early, "bad" deferred, with priority queues keyed by tiered accessibility and parcel-size guards. Third, we assign FAR via an accessibility-weighted linear model (anchored at a baseline $\bar f$) and convert to heights using footprint and storey factors. We report achieved shares and construction-share deviation, and explore policy space via LHS sampling, non-dominated sorting, and knee-point selection on the Pareto front.

\subsection{Accessibility analysis}\label{sec:access}
\textbf{Graph construction and analysis radius.} Given a street network, we clean and merge it, then build a space-syntax graph and compute centralities. Formally, let the street network be a graph $G=(V,E$ where each edge or segment becomes an analyzable element. For any pair of edges $p$, $q$, the path cost $d(p,q)$ depends on the selected metric and is evaluated only on paths of length $\leq r$ (with $r=\infty$ for city level), where $r$ is the specified radius.\\

\noindent \textbf{Segment-level accessibility metrics.} We compute four families of edge scores: Metric Choice (betweenness), Metric Integration (closeness), Angular Choice (betweenness under angular cost), and Angular Integration (closeness under angular cost):
\begin{itemize}
    \item Choice (betweenness), which counts the number of times each segment falls on the shortest path between all pairs of segments within the given distance (radius $r$). The calculation of Metric Choice and Angular Choice is similar:
    \begin{equation}
        \text{Choice}(e) = \sum_{s \neq t \neq e} \frac{\sigma_{s,t}(e)}{\sigma_{s,t}}, \text{with } d(s,t) \leq r,
    \end{equation}
    where $\sigma_{s,t}(e)$ is the number of shortest paths between nodes $s$ and $t$ (according to some metric, such as metric distance and angular distance) that go through edge $e$. If it is Angular Choice, shortest paths use angular distance (sum of turning angles) to identify ``the straightest routes''. And $\sigma_{st}$ is the total number of shortest paths between $s$ and $t$. The summation $\sum_{s \neq t \neq e}$ means we add this fraction over all distinct pairs of nodes $s$ and $t$, excluding edge $e$ itself.  
    \item Integration (closeness), which measures how close each segment is to all others within the given radius:
    \begin{equation}
        \text{Integration}(i) = \frac{1}{\sum_{j \neq i} d(i, j)} \quad \text{with } d(i, j) \leq r,
    \end{equation}
    where $i$ is the node we are evaluating and $j$ is another node in the network. $d(i, j)$ is the shortest path distance between nodes $i$ and $j$, measured with respect to metric length or angular distance. $\sum_{j \neq i} d(i, j)$ is the total "effort" needed for node $i$ to reach all the other nodes $j$. If it is angular distance, we will substitute the calculation of total effort with $\text{NTD}_{\text{ang}}(i)$, which is the normalized total angular distance from node $i$ to all other nodes.
\end{itemize}

\noindent \textbf{From segment scores to block accessibility.} For each block polygon, we store the maximum nearby or adjacent street segment scores as the block’s accessibility. We can write this aggregation explicitly as $A_i = \max_{e \in \mathcal{N}(i)} S_e$, where $\mathcal{N}(i)$ is the set of road segments intersecting/touching block $i$, $S_e$ is the selected segment centrality. The score of each segment can be a weighted sum of multiple centrality calculations. Formally, $S_e = \sigma S^{choice}_e + (1-\sigma)S^{Integration}_e$, where $\sigma \in [0,1]$. \YS{This methodological choice is grounded in the "movement economy" theory of space syntax, which posits that land-use potential—particularly for retail and commercial density—is driven by the highest tier of movement available to the plot. In urban real estate economics, the development capacity of a parcel is typically determined by its "best" or "active" frontage (e.g., the commercial high street), rather than an average of all bounding edges, which would incorrectly penalize blocks with necessary but low-integration service alleys \citep{hillier2007space, porta2006network}. We acknowledge that the limitations of using the maximum value as the block-level aggregator may being overstating accessibility for blocks with one high-access edge and several low-access edges since lesser-connected sides are ignored, which may be an inherent source of uncertainty in our segment-block translation. Alternatives, such as a mean or frontage-length-weighted average, could be explored in future extensions but may dilute peak signals critical for activity-seeking uses like retail. It causes the algorithm to under-predict the suitability of prime corner lots or arterial-facing blocks, leading to a "salt-and-pepper" distribution of land uses that fails to form the continuous commercial corridors observed in real cities. For less frontage-dependent uses (e.g., residential or industrial), this approach may amplify extremes. For instance, residential exclusivity often correlates with segregation or quietness. However, our priority queues and tiered allocation introduced in Section \ref{sec:allocation} temper such effects by sequencing allocations where residential and industrial uses are deferred to lower-priority placements.}\\

\noindent \textbf{Multi-scale accessibility.}\label{multi_scale_acc} We specified a five-tier urban service hierarchy: community, community cluster, life circle, district, and city. Each level is associated with a corresponding service radius. By default, the range of living radius within a community is 200-400 meters, community cluster 600-900 meters, life circle ($\approx$15-minute walk) 1200-1500 meters, district 2-3 kilometers, and at the city scale, we use 10-15 kilometers. These ranges are not hard-set and can be set explicitly implemented in the analysis setup. Note that not every city includes all these hierarchies. The number of tiers is determined by the number of input radii. Then we compute centralities at each radius and map the results to blocks. So every block will be assigned the same number of accessibility scores as the number of radii. Formally, if we compute accessibility at $L$ radii $\left\{ r_0, ..., r_{L-1}\right\}$ (e.g., community to city), the accessibility of block $i$ at level $l$ is $A_{i,l} = A_{i}(r_l)$, the result will be normalized at each level for the downstream land-use allocation.\\

\subsection{Land-use allocation}\label{sec:allocation}
\noindent \textbf{Inputs and setup.} Given a fixed block set $\mathcal{B}$, and a per-tier accessibility vector $\mathbf{A}_{i\in\mathcal{B}}=\big(A_{i,\ell}\big)_{\ell\in\mathcal{L}}$, where $\mathcal{L}$ is the set of active tiers, we assign each block a dominant use from $\mathcal{U}=\{R,A,G,B,I,T,E,F\}$ (i.e., Residential, Admin/Office, Green, Business/Retail, Industrial, Transport, Education, Food). The goal is to match the site-wide target land-use percentages vector $\mathbf{s}^*=(s_u^*)_{u\in\mathcal{U}}$, where $\sum_u s_u^*=1$. Each block $i$ has lot area $Area^{lot}_{i}$. This implies the total target area of land-use type $u$ is $R_u=s_u^*\cdot Area^{\mathrm{total}}$, where $Area^{\mathrm{total}}=\sum_i Area^{lot}_{i}$. \YS{Additionally, to operationalize the hierarchy of these service tiers, we apply a Rank Sum weighting framework,  a standard technique in Multi-Criteria Decision Analysis (MCDA) for converting ordinal planning priorities into cardinal allocation weights \citep{LakmayerDanielson2025EfficientWeightRanking, Sureeyatanapas2016Comparison}. We posit a linear hierarchy where the "City" scale is the primary driver of structure, followed sequentially by finer scales. We assign rank scores $w_l$ linearly: the City tier receives a score of 12, the District a score of 11, down to the Community tier at 8 (i.e., $w_l \in \{$city - 12, district - 11, life circle - 10, community cluster - 9, community - 8$\}$). These descending cardinal weights serve as a heuristic rooted in planning hierarchies (e.g., central place theory \citep{getis1966christaller} and 15-minute city \citep{smartcities4010006}), prioritizing coarser scales to anchor strategic uses before local ones, as larger radii capture systemic connectivity \citep{hillier1976space}. While the specific integers are tunable, their linear spacing reflects the Rank Sum assumption that the explanatory power of accessibility tiers decays linearly rather than exponentially or logarithmically. While not empirically calibrated here, as our emphasis is on the overall framework, future extensions could explore variations for context-specific tuning. The normalized weight for tier $\ell$ is computed as $\text{LevelPct}(\ell)=\frac{w_l}{\sum_{k \in \mathcal{L}}w_k}$. For example, if there are only three active tiers (e.g., $\mathcal{L}=\{$district, community cluster, community$\}$) in the use case, the total land-use area of type $u$ at the community cluster tier $cs$ is $R_{u, cs}=\frac{9}{11+9+8} \cdot R_u=0.32R_u$. The total area of type $u$ being assigned at tier $\ell$ is $R_{u,\ell} = \text{LevelPct}(\ell)\cdot R_u = \text{LevelPct}(\ell)\cdot s^*_{u} \cdot Area^{total}$. Then we assign a land-use priority to reflect the urban planner's focus and intention. For example, if it is a pro-business planning scheme, then we can prioritize business land use and set a strict sequence order $\pi$ over $\mathcal{U}$ (e.g., $B\!\succ\!A\!\succ\!E\!\succ\!F\!\succ\!G\!\succ\!R\!\succ\!I\!\succ\!T$). The arrangement guarantees blocks with the best accessibility scores are of business land-use. Furthermore, the model automatically splits the input land-use list into ``good'' uses ($\{A,G,B,E,F\}$) and ``bad'' uses ($\{I, T\}$). Residential land-use is treated separately. Finally, a monotonic list of distance thresholds $\mathbf{C}=\{C_0<C_1<\dots<C_{L-1}\}$ that defines the multi-level clustering is needed. At each level, the threshold should be compatible with the tiers and the range of living radii specified in the multi-scale accessibility calculation.}\\

\noindent \textbf{Tier constraints and "bad" land uses.} Two constraints are enforced by level: (1) assignment proceeds from coarse tiers (e.g., city or district) down to community level; (2)``Industrial'' and ``Transportation'' land uses ($I,T$) are only held for higher tiers (city and district scale) and placed late. These rules ensure early placement of activity-seeking land-uses (i.e., $G,B,F,E,A$) being pushed onto accessible tiers first and $I,T$ being deferred to coarser tiers. Meanwhile, they steer industrial and transportation to a larger scale where impacts are easier to absorb before local services and activity-seeking land-uses are distributed.\\

\noindent \textbf{Hierarchical clustering of blocks based on accessibility.} We implement hierarchical clustering over blocks to turn the accessibility score field into service basins at successive living-radius tiers. Specifically, at each travel scale, we pick local centers where the score is high so that they are good places to host services. Then, for each center, we collect the blocks within the radius of the corresponding tier. The set of groups at the tier is denoted as $\mathcal{C}_\ell$. A group is a service basin or catchment. Doing this at several travel distances creates nested groups. Because groups are nested across scales, the allocator can move top-down: place higher-priority uses (defined by $\pi$) in the most connected groups first, then fill others. The model scans queues inside groups (not the whole site), so that land-use allocation can be performed in parallel across clusters to speed up the computation. And the formed clusters will also be handed off to the FAR calculation, where the same accessibility scores that formed the groups also guides intensity, keeping land-use and heights consistent.\\

\noindent \textbf{Land use assignment from highest tier down.} Based on the case in "Inputs and setup", there are three active tiers specified. Within each level, the model does:
\begin{itemize}
    \item Step one: The model will compute how much non-residential area to allocate at this tier. At the top tier, it includes all "bad" uses $\mathcal{U}_{\mathrm{bad}}=\{I,T\}$ plus a share of "good" uses $\mathcal{U}_{\mathrm{good}}=\{G,B,F,E,A\}$ weighted by the tier weight. At lower tiers, it allocates only the "good" uses weighted by the tier weight. Formally, the total non-residential area at the chosen level $\ell$ to allocate is $R_{\ell,\mathcal{U}_{\mathrm{good}}}= \text{LevelPct}(\ell)\cdot Area^{total} \cdot \sum_{u\in\mathcal{U}_{\mathrm{good}}} s_u^*$.
    \item Step two: The model chooses which clusters are eligible at this tier. A cluster is available only if the sum of remaining area in its blocks is $\leq 60\%$ of that cluster’s total area. \YS{We define this 60\% threshold as the Cluster Integrity Threshold ($\tau_{int}=0.6$). It is a tunable constraint informed by site percolation thresholds in spatial networks, which suggest that spatial connectivity and the potential for agglomeration break down significantly when occupancy (or availability) drops below $\approx 59\%$ \citep{ZhangZhu2013SpatialMultiresolutionClusterDetection, KennedyWilkinsonBalch2003ConservationThresholds}. In urban design terms, this prevents the "Swiss cheese" effect, where high-priority uses are forced into fragmented, non-contiguous residuals of a service basin. By enforcing $\tau_{int}$ to be around 50\% to 70\% retention in subdivision rules, the model guarantees that any selected service basin retains sufficient contiguous mass to support a cohesive functional zone. This parameter is exposed as a tunable variable: for infill scenarios where fragmentation is unavoidable, planners may lower $\tau_{int}$ to accept greater fragmentation.}    
    \item Step three: Split the tier’s service area across the available clusters by cluster area share. Within each chosen cluster, compute the land-use targets for this tier. At the top level, the model will initially carve out the total land area for $I$ and $T$. If the global target percentage share for $\mathcal{U}_{\mathrm{bad}}$ is $s^*_{\mathcal{U}_{\mathrm{bad}}}=s^*_{I}+s^*_{T}$, then area  $R_{\mathcal{U}_{\mathrm{bad}}}=s^*_{\mathcal{U}_{\mathrm{bad}}} \cdot Area^{total}$ will be carved out at the highest tier. The model distributes the remainder among the "good" uses proportionally to their global shares (excluding R/I/T). At non-top tiers, it only splits among the "good" uses.
    \item Step four: Assign those land-use targets to actual blocks using a priority queue. The model builds a Priority Queue (PQ) of the cluster’s blocks that still have remaining area, keyed by the block’s accessibility at this tier (i.e., $A_{i,\ell}$). It then allocates each target land-use to blocks by repeatedly popping from the PQ, assigning area, and reinserting the block if it still has remaining area. For "good" uses, it uses the accessibility order as-is. At the top tier, it inverts the queue and assigns "bad" uses to the least accessible locations first.
    \item Step five: Respect minimum parcel sizes while assigning. Each tier has a minimum leftover area threshold (City/District - $2500m^2$, living circle - $2000m^2$, community cluster - $1000 m^2$, community - $500 m^2$). When giving area to a land-use, if finishing the remaining target would leave the block with less than the tier’s minimum, the model gives the entire remaining block to that land-use to avoid slivers.
\end{itemize}

\noindent \textbf{After all tiers.} The remaining land becomes Residential ($R$). For each block, any area not assigned to non-$R$ uses is converted to $R$. If a block ended up with multiple uses, it is split geometrically by the use ratios. Otherwise, its single dominant use is set as the block’s land-use. The model outputs:
(1) split geometries and dominant-use label $u(i)$ for each block (or per-block shares in the mixed variant),
(2) achieved shares $\hat s_u$, 
(3) the carried-through accessibility tensor, which is reused to compute FAR weights in Section~\ref{sec:FAR}. Program assignment and area accounting are then mapped to FAR/height in the next stage, making consistent the practice of turning program area into height via buildable area. We demonstrate the pseudo-code for the land use assignment process as shown in Algo.\ref{algo:cluster}.\\

\begin{algorithm}
\scriptsize
\caption{Greedy priority-guided land use assignment}\label{algo:cluster}
\begin{algorithmic}[1]
\Require Blocks $i\in\mathcal{B}$ with lot area $Area^{lot}_{i}$ and tiered accessibility $A_{i,\ell}$; \quad tiers $\ell\in\mathcal{L}$ ordered coarse$\rightarrow$fine; \quad for each tier $\ell$, a partition of blocks into clusters $\mathcal{C}_\ell=\{c\}$; \quad global target shares $s_u^*$ for uses $u\in\mathcal{U}$; \quad priority order $\pi$; \quad tier weights $w_\ell$ with $\mathrm{LevelPct}(\ell)=w_\ell/\sum_{k\in\mathcal{L}}w_k$; \quad bad uses $\mathcal{U}_{\mathrm{bad}}=\{\mathrm{I},\mathrm{T}\}$; \quad good uses $\mathcal{U}_{\mathrm{good}}=\mathcal{U}\setminus(\mathcal{U}_{\mathrm{bad}}\cup\{\mathrm{R}\})$; \quad min-parcel by tier $m_\ell$.

\State $Area^{total}\leftarrow \sum_{i\in\mathcal{B}}Area^{lot}_{i}$;\quad $\textstyle S_{\mathrm{good}}\leftarrow\sum_{u\in\mathcal{U}_{\mathrm{good}}} s_u^*$;\quad $S_{\mathrm{bad}}\leftarrow \sum_{u\in\mathcal{U}_{\mathrm{bad}}} s_u^*$.
\State Initialize remaining area on each block: $R_i\leftarrow Area^{\mathrm{lot}}_i$. Initialize assigned matrix $x_{i,u}\leftarrow 0$.
\For{$\ell$ in tier order (coarse $\rightarrow$ fine)}
  \State \textbf{Tier pools:}
  \If{$\ell$ is top tier}
     \State Nonresidential pool $P_\ell \leftarrow (S_{\mathrm{bad}} + S_{\mathrm{good}}\cdot \mathrm{LevelPct}(\ell))\cdot Area^{\mathrm{total}}$.
  \Else
     \State $P_\ell \leftarrow (S_{\mathrm{good}}\cdot \mathrm{LevelPct}(\ell))\cdot Area^{\mathrm{total}}$.
  \EndIf
  \State \textbf{Eligible clusters:} $\mathcal{C}_\ell^{\mathrm{avail}}\leftarrow \{\,c\in\mathcal{C}_\ell:\sum_{i\in c}R_i \ge 0.6\cdot \sum_{i\in c}Area^{\mathrm{lot}}_i\,\}$.
  \State \textbf{Area shares by cluster:} For each $c\in\mathcal{C}_\ell^{\mathrm{avail}}$,
        $\alpha_c \leftarrow \dfrac{\sum_{i\in c}Area^{\mathrm{lot}}_i}{\sum_{c'\in\mathcal{C}_\ell^{\mathrm{avail}}}\sum_{j\in c'}Area^{\mathrm{lot}}_j}$.
  \State \textbf{Per-cluster, per-use targets $T_{u,\ell,c}$:}
  \For{each $c\in\mathcal{C}_\ell^{\mathrm{avail}}$}
     \If{$\ell$ is top tier}
        \State For $u\in\mathcal{U}_{\mathrm{bad}}$: $T_{u,\ell,c}\leftarrow \alpha_c\,(s_u^*\cdot Area^{\mathrm{total}})$.
        \State For $u\in\mathcal{U}_{\mathrm{good}}$: $T_{u,\ell,c}\leftarrow \alpha_c\,(s_u^*\cdot Area^{\mathrm{total}}\cdot \mathrm{LevelPct}(\ell))$.
     \Else
        \State For $u\in\mathcal{U}_{\mathrm{good}}$: $T_{u,\ell,c}\leftarrow \alpha_c\,(s_u^*\cdot Area^{\mathrm{total}}\cdot \mathrm{LevelPct}(\ell))$.
        \State For $u\in\mathcal{U}_{\mathrm{bad}}$: $T_{u,\ell,c}\leftarrow 0$.
     \EndIf
  \EndFor
  \State \textbf{Assign to blocks via accessibility PQ:}
  \For{each $c\in\mathcal{C}_\ell^{\mathrm{avail}}$}
     \For{$u$ in priority order $\pi$ with $T_{u,\ell,c}>0$}
        \State Build priority queue PQ with all $i\in c$ having $R_i>0$, keyed by $A_{i,\ell}$;
               use max-heap if $u\in\mathcal{U}_{\mathrm{good}}$, min-heap if $u\in\mathcal{U}_{\mathrm{bad}}$ (only possible at top tier).
        \While{$T_{u,\ell,c}>0$ and PQ not empty}
           \State $i\leftarrow \text{pop}(\text{PQ})$.
           \State Candidate give: $g\leftarrow \min\{R_i,\;T_{u,\ell,c}\}$.
           \If{$R_i-g < m_\ell$ \textbf{and} $R_i \ge m_\ell$} \State $g\leftarrow R_i$ \Comment{avoid sub-minimum leftovers} \EndIf
           \State $x_{i,u}\leftarrow x_{i,u}+g$;\quad $R_i\leftarrow R_i-g$;\quad $T_{u,\ell,c}\leftarrow T_{u,\ell,c}-g$.
           \If{$R_i>0$} \State $\text{push}(\text{PQ}, i)$ \EndIf
        \EndWhile
     \EndFor
  \EndFor
\EndFor
\State \textbf{Residual to Residential:} For all $i$, set $x_{i,\mathrm{R}}\leftarrow x_{i,\mathrm{R}}+R_i$ and $R_i\leftarrow 0$.
\State \textbf{Post-process geometry:} For any block with multiple positive $x_{i,u}$, split the polygon by ratios $x_{i,u}/Area^{\mathrm{lot}}_i$; else label it by its single use.
\end{algorithmic}
\end{algorithm}

\noindent \textbf{Achieved shares and diagnostics.} After the assignment, we report achieved shares as the below:
\[
\hat s_u \;=\; \frac{1}{Area^{\mathrm{total}}}\sum_{i:\,u(i)=u}Area_i^{\mathrm{lot}},\qquad \sum_u \hat s_u = 1.
\]
And compute the target deviation $D_{LU}=\sum_u(\hat s_u - s_u^*)^2$ used as one objective function.

\subsection{FAR and height calculation}\label{sec:FAR}
\noindent \textbf{Inputs and setup.} Given the split block geometries $i \in \mathcal{S}$ and land-use assignment results ${u(i)}$, the goal is to assign each lot a floor-area ratio (FAR) and implied building height. We first need to specify a global building area target $B^{total}$. Then each land-use type $u$ has a global building area target $B_u = \gamma^*_u \cdot B^{\mathrm{total}}$, where $\gamma^*_u$ is the construction share ratio (CSR) for land-use type $u$. Similar to land-use allocation, we need to input a site-wide target CRS vector $\mathbf{\gamma}^*=(\gamma^*_u)_{u\in\mathcal{U}}$, where $\sum_u \gamma^*_u=1$. The model also receives normalized accessibility scores $A_{i,\ell}$ for each lot at each living-radius tier $\ell$ from the output of Section \ref{multi_scale_acc}. Then we input an accessibility weight vector $\rho^*=(\rho_{\ell})_{\ell \in \mathcal{L}}$ whose dimension is the number of active tiers. We will compute the weighted sum of accessibility scores for each lot. Different from the normalized weights $\text{Level}(\ell)$ for active tiers, the weight vector $\rho^*$ is used to express the planning intention. For instance, the community tier receiving higher weight than others emphasizes the importance of local accessibility over long-distance reach. The weighted sum of accessibility scores for each lot becomes the key explanatory variable for intensity allocation. Formally, the weighted accessibility score of lot $i$ is 
\[
\tilde{A}_i = \sum_{\ell\in\mathcal{L}} \rho_\ell\, A_{i,\ell}, 
\qquad \text{with } \sum_{\ell\in\mathcal{L}} \rho_\ell = 1.
\]\\

\noindent \textbf{Accessibility-weighted FAR model.} Within each land-use type, the model distributes the target building area proportionally to block accessibility. Let $Area^{lot}_i$ be the lot area of lot $i$ and $\tilde{A}_i$ its accessibility score. For a set of lots $\mathcal{S}$ with target total building area $B^{total}$, the algorithm solves a linear form
\[
FAR_i = \alpha \cdot \tilde{A}_i + \beta,
\]
with coefficients chosen so that two conditions hold: (1) The weighted sum of building area matches the target. That is $\sum_{i \in \mathcal{S}} FAR_i \cdot Area^{lot}_i = B^{total}$. (2) The average FAR across block centers near a reference value (here we specify $\bar f = 0.8$) to avoid negative or extreme values. The reference average FAR $\bar f$ as an anchor is fixing the condition where, when a block has the worst accessibility, its FAR should be around a modest baseline, not negative or extreme. The two constraints are achieved by setting slope $\alpha$ and intercept $\beta$ from block statistics:
\[
\alpha = \max\!\left(
  \frac{\bar f - \dfrac{B^{total}}{\sum_i Area^{lot}_i}}
       {\min(\tilde{A}_i) - \dfrac{\sum_i \tilde{A}_i Area^{lot}_i}{\sum_i Area^{lot}_i}},
  \; 0
\right),
\qquad
\beta = \frac{B^{total}}{\sum_i Area^{lot}_i}
        - \frac{\sum_i \tilde{A}_i Area^{total}_i}{\sum_i Area^{total}_i}\,\alpha,
\]
where $\bar f - \dfrac{B^{total}}{\sum_i Area^{lot}_i}$ is the gap between the chosen baseline $\bar f$ and the average FAR implied by the global target $B^{total}$. And $\min(\tilde{A}_i) - \dfrac{\sum_i \tilde{A}_i Area^{lot}_i}{\sum_i Area^{lot}_i}$ is the gap between the least accessible lot’s score and the average accessibility weighted by lot area. Dividing the two gives a slope $\alpha$ that adjusts FAR so that, when a block has the lowest accessibility, its FAR tends toward $\bar f$, and when a block has average accessibility, FAR averages to the global target. $\beta$ shifts the whole line so the global building-area constraint is met. Thus $FAR_i = \alpha \cdot \tilde{A}_i + \beta$ ensures that higher-accessibility blocks are given higher intensity while total building area is respected.\\

\noindent \textbf{Height estimation.} Once FAR values are assigned, heights are derived via $H_i = \frac{FAR_i \cdot Area^{lot}_i}{Area_i^{\mathrm{footprint}}} \cdot \phi$, where $Area_i^{\mathrm{footprint}}$ is the typical buildable footprint and $\phi$ a storey height (e.g., 3.6 m/floor). This step turns the program area into physical heights consistently with land-use. \YS{For simplicity, we applied a uniform typical footprint ratio across all uses in this study. This turns the program area into physical height but is a coarse approximation: in reality, building coverage (footprint/lot area) varies by land use (e.g., residential buildings might cover 20–35\% of a lot, whereas commercial buildings often cover 30–50\%). Our generic assumption ignores these differences, so it may overestimate heights for uses that normally have lower lot coverage and vice versa. In practice, one could introduce use-specific coverage parameters to refine the FAR-to-height conversion for each land-use type – an enhancement to improve realism. Here, we use a single typical footprint as a tractable simplification to close the model loop.}\\

\noindent \textbf{Output and diagnostics.} Overall, the FAR allocator outputs (1) block-level FAR values ${FAR_i}$, (2) implied heights ${H_i}$, and (3) diagnostics of achieved building areas per land-use. This ensures consistency: land-use determines “what” is built, while accessibility-driven FAR allocation determines “how much” and “how tall”. Specifically, we verify whether the global building area targets $B_u$ for each land-use $u$ are met. For each land-use type, the achieved building area is computed as 
\[
\hat{B}_u = \sum_{i:\,u(i)=u} \mathrm{FAR}_i\, Area_i,
\]
where $A_i$ is the lot area of block $i$. The achieved share of the construction area is then
\[
\hat{\gamma}_u = \frac{\hat{B}_u}{\sum_i \mathrm{FAR}_i\, Area_i},
\qquad
\sum_u \hat{\gamma}_u = 1.
\]
We define the deviation from targets as
\[
D_B =  (B^{total} - \sum_{u\in\mathcal{U}} \hat{B}_u)^2, \quad
D_{CS} = \sum_{u\in\mathcal{U}} \left(\hat{\gamma}_u - \gamma^*_u \right)^2,
\]
which measures how closely the achieved total construction area approaches the target and how closely the allocation matches the input construction share ratios $\gamma^*$. In practice, small deviations may occur due to minimum-parcel constraints, numerical rounding, or accessibility weighting. These diagnostics allow inspection of whether intensities are consistent with design intentions.\\

\subsection{Optimization for Pareto solutions}
\noindent \textbf{Decision space.} We search over a compact policy space $\Theta$ whose components are: analysis radii (tier set $\mathcal{L}$ and numeric radii), land-use share vector $s^*$, land-use priority order $\pi$, construction share vector $\gamma^*$, target global building area $B^{total}$ and accessibility weights $\rho$.\\

\noindent \textbf{Sampling and evaluation.} We generate $N$ design points ${\theta^{(k)}}_{k=1}^N\subset\Theta$ via Latin Hypercube Sampling (LHS) to cover the policy space efficiently. For each $\theta^{(k)}$, we run the proposed pipeline and compute the following objective metrics:
\[
\bigl(AU^{(k)},\; D^{(k)}_{LU},\; D^{(k)}_{CS},\; D^{(k)}_B, \; JH^{(k)}_{\mathrm{pen}}\bigr).
\]
To avoid the impact of different degrees of magnitude and the assumption of the distributions of these metrics, we normalize these metrics by $m_k = \frac{v_k-v_{min}}{v_{max}-v_{min}}$, where $m_k$ is the normalized value of sample $k$, and $v_k$, $v_{min}$, $v_{max}$ are respectively the original value of sample $k$, minimum and maximum values across the chosen metric type. Additionally, since $D^{(k)}_{LU},\; D^{(k)}_{CS},\; D^{(k)}_B$ represent deviations from the targets, we combine these three metrics and normalize the result so that $D^{(k)}_{total} \in [0,1]$ expresses the total deviation from the target by sample $k$. Therefore, the objective metrics for each $\theta^{(k)}$ is three dimensional and is easier for visualization:
\[
\bigl(AU^{(k)},\; D^{(k)}_{total}, \; JH^{(k)}_{\mathrm{pen}}\bigr).
\]\\

\noindent \textbf{Objectives.} We evaluate the three planning objectives below, where all are minimized except $AU$. For the convenience of finding Pareto solutions, we modify the values of $AU$ to $1-AU$, inverting maxima and minima, and normalizing the value range to that of the other objectives. Therefore, all objectives are minimized and within [$0,1$].
\begin{itemize}
    \item Accessibility utility. We reward placing intensity where it is most reachable:
    \[
        min \; AU = 1-\frac{\sum_i \mathbf{1}[\,u(i)\in \mathcal{U}_{good} \cup \{R\}\,] \cdot \mathrm{FAR}_i \cdot Area_i^{lot} \cdot \tilde{A}_i}
                  {\sum_i \mathrm{FAR}_i \cdot Area_i^{lot}}
             \in [0,1],
    \]
    where $\tilde A_i=\sum_{\ell \in \mathcal{L}} \rho_\ell \cdot A_{i,\ell}$. This is a FAR- and area-weighted accessibility index emphasizing activity-seeking uses.
    \item Target deviation of land-use and construction shares:
    \[
        \; D_B =  (B^{total} - \sum_{u\in\mathcal{U}} \hat{B}_u)^2, \quad
        \; D_{LU} = \sum_{u\in\mathcal{U}}\big( \hat s_u - s_u^* \big)^2, \quad 
        \; D_{CS} = \sum_{u\in\mathcal{U}} \left(\hat{\gamma}_u - \gamma^*_u \right)^2,
    \]
    with $\hat s_u$ from Section~\ref{sec:allocation}, $\hat{B}_u$ and $\hat\gamma_u$ from Section~\ref{sec:FAR}. And the total deviation for one sample is $D_{total} = D^{norm}_B + D^{norm}_{LU} + D^{norm}_{CS}$, and the objective is
    \[
        min \; D^{norm}_{total} = \frac{D_{total} - D^{total}_{min}}{D^{total}_{max} - D^{total}_{min}} \in [0,1].
    \]
    \item Jobs–housing balance. With $\text{Jobs} \approx Area_B+ Area_F$ and $\text{Housing} \approx Area_R$:
    \[
        min \; JH_{penalty} \;=\; \left(\frac{\text{Jobs}}{\text{Housing}}-r_0 \right)^2,
    \]
    where $r_0$ is a policy target. Here, we specify $r_0=1.2$ aligning with the conclusions by Peng et. al. \citep{Peng1997JobsHousing} and Xu et. al. \citep{XU2017126} suggesting 1.2 is a tipping point below which commuting increases more sharply. The penalty will also be min-max normalized. \YS{The formulation uses a coarse area-based proxy for jobs (Business + Food) and housing (Residential), acceptable for high-level modeling but simplifying actual densities - in reality, jobs or dwellings per unit area vary by development type. Nonetheless, it provides a tractable indicator of jobs–housing balance.}
\end{itemize}

\begin{figure}[H]
\begin{center}
\includegraphics[angle=0,
scale=0.105,
trim={0cm 0cm 0cm 0cm},clip]{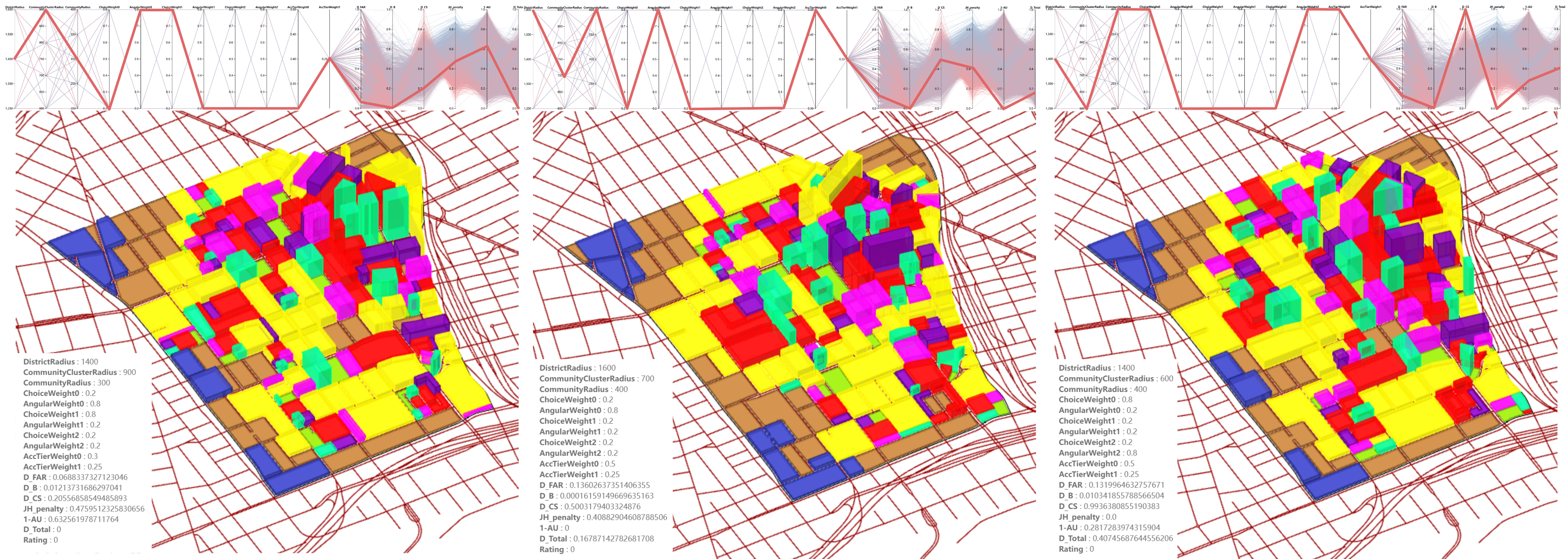}
\caption{Three instances of Pareto policies by respectively selecting the least $D_{Total}$ (left), $1-AU$ (middle), and $JH_{penalty}$ (right). Each polyline across the parallel coordinates indicates one solution. The corresponding polylines of the selected optimal solutions are highlighted.} 
\label{fig:Parallel_coordinates}
\end{center}
\end{figure}

\noindent \textbf{Pareto set extraction and knee selection.} From the evaluated set, we perform non-dominated sorting to obtain the Pareto front $\mathcal{P}$ (the set of non-dominated solutions). For reporting a single recommended plan, we identify a knee point $\theta^\dagger\in\mathcal{P}$ by maximizing normalized curvature of the front (or, equivalently, minimizing the distance to the utopia point after objective-wise min–max normalization). Since we are seeking minimum of each objective, the utopia point is naturally the origin. If two solutions are equally close to the utopia point, we prefer the one with lower $D_{CS}$ (i.e., better match to construction share targets), and if there is still a tie, then lower $D_{LU}$ (better land-use share match). These tie-breakers align with regulatory or policy priorities: first intensity consistency, then land-use mix consistency. Overall, we return (1) the Pareto set $\mathcal{P}$ with full objective vectors; (2) per-solution summaries (achieved shares, $AU$, $JH_{penalty}$, height statistics); and (3) the knee solution $\theta^\dagger$ with its full spatial outputs (maps of $u(i)$, $FAR_i$, $H_i$). This enables planners to compare trade-offs transparently—e.g., how much $AU$ improvement is “worth” a small increase in $JH_{penalty}$—and to select policies aligned with local priorities as shown in Fig~\ref{fig:Parallel_coordinates}.\\

\section{Case Study}\label{sec:case}
\subsection{Site and Experimental setup}
The selected site is around $2.5km^2$ in West Oakland, California, where long-standing residential blocks to the east and north meet heavy industrial and seaport/rail logistics to the west. Major commercial corridors line on the south boundary and northeast corner (Fig. \ref{fig:current_site_land_use}a). Since the testbed is at the junction of diverse land uses, we also conduct a reality-check evaluation with minimum inputs: beyond optimization and Pareto exploration, we compare the model’s dominant-use assignment to observed land use at the block level, reporting overall agreement, class-wise precision/recall, and share deviation. This validates whether the accessibility-guided allocator reproduces West Oakland’s existing pattern before we use it for counterfactual policy design.\\

\noindent We obtained the street network from OpenStreetMap \citep{osm} and generated block polygons with buildable lot area. The site's current zoning map is cross-walked to our model classes $\mathcal{U}=\{R,A,G,B,I,T,E,F\}$ using the rule $RM \rightarrow R$, $OS \rightarrow G$, $CC \rightarrow B+F \; (\text{split 3/1})$, $RU \rightarrow R+I \; (\text{split 50/50})$, $HBX \rightarrow A+R \; (\text{split 50/50})$, $CIX \rightarrow I+T (\text{split 50/50})$, and $U \rightarrow E$. Area shares are computed inside the study boundary by color segmentation of the map. The resulting baseline vector is
\[
    s^{obs} = (F:5.8\%, B:17.4\%, E:4.7\%, G:6.5\%, A:5.9\%, R:33.3\%, I:19\%, T:7.4\%).
\]
The observed shares $s^{obs}$ serve both as calibration targets for the reality-check and as starting points for policy sampling of optimization. We set LHS bounds for all decision variables. For example, three tiers of radii for accessibility analysis is specified: $r_{district} \in [1200m, 1600m]$, $r_{community \; cluster} \in [600m, 900m]$, $r_{community} \in [200m, 400m]$. The weight to average Choice and Integration for each street segment $\sigma \in [0,1]$. And the value range of each element of the accessibility weight vector $\rho_\ell \in [0,1]$ to reflect the emphasis for each tier $\ell \in \mathcal{L}$ when calculating FAR. We specified that the total construction target is $6km^2$. Furthermore, to limit the size of policy space, the share of construction area in each land use is fixed as 
\[
\gamma^* = (A:10\%,B:25\%,G:0,I:5\%,T:5\%,R:35\%, E:10\%,F:10\%).
\]
Altogether, we drew around $7700$ samples and computed Pareto sets.

\subsection{Results}
\begin{figure}[H]
\begin{center}
\includegraphics[angle=0,
scale=0.16,
trim={0cm 0cm 0cm 0cm},clip]{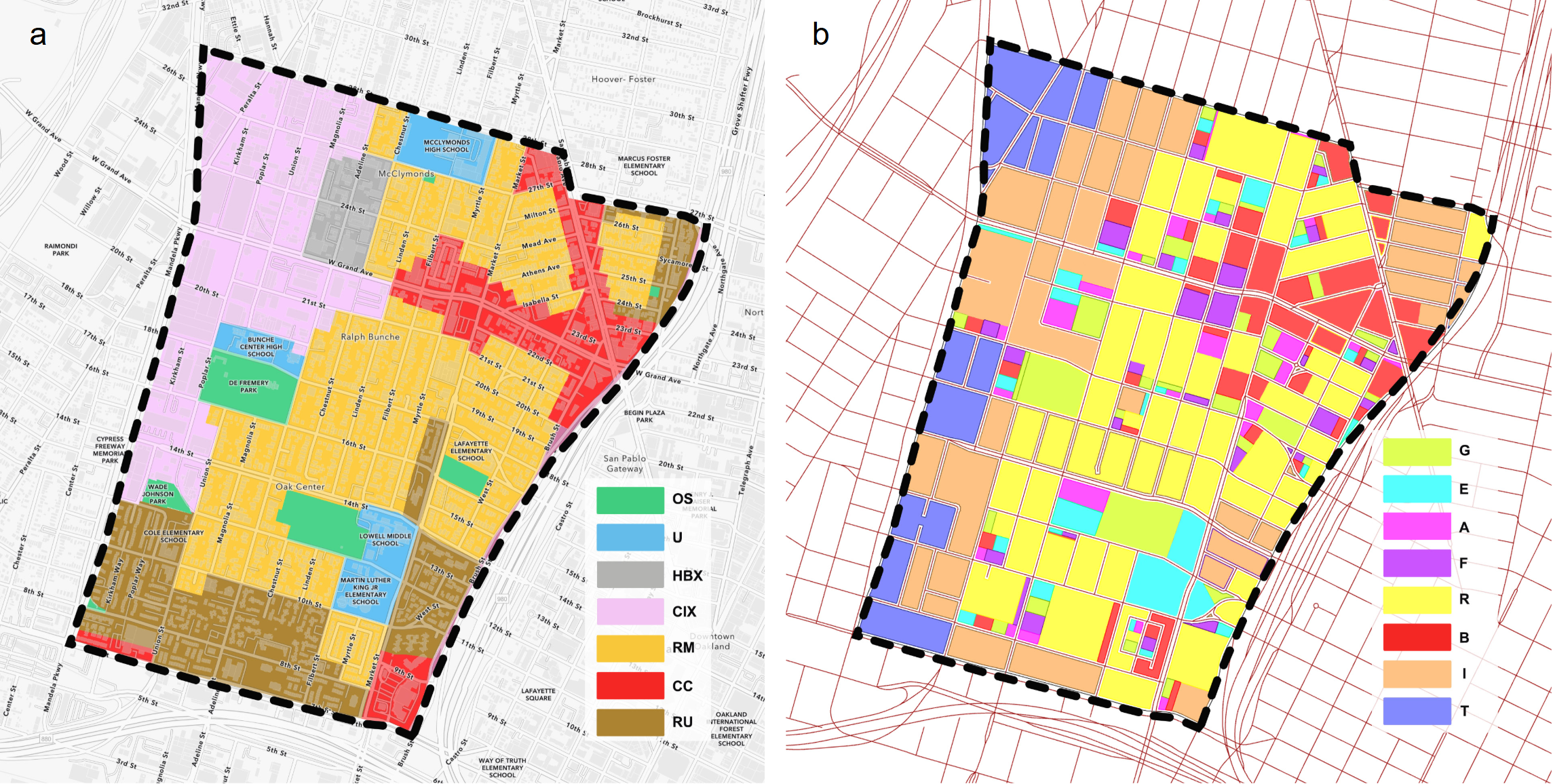}
\caption{Comparison of observed land use and generated land use map with minimum inputs. a) Current land use map used for validation. It is the snapshot from the interactive zoning map of City of Oakland, California \citep{oakland_experiencebuilder}. \YS{The legend is zoning code. OS: open space; U:education; HBX: housing and business mix; CIX: industrial and business mix; RM: mixed housing type residential; CC: community commercial; RU: urban residential.} b) Baseline simulation result using land use percentages share vector $s^{obs}$ retrieved by color segmentation analysis from Plot a. The land use sequence order specified in $s^{obs}$ is also applied. \YS{The legend follows our land use class. G: open space; E: education; A: administration; F: restaurant; R: residential; B: business; I: business and industrial mix; T: transportation. Note that, in plot a, we notice there are transportation facilities in both southwest (Oakland Bart station) and northwest (bus station) corner but are categorized as community commercial and industrial/business mix. Therefore, we add transportation land use in our simulation. Its target land use share can be set to 0 if necessary.}} 
\label{fig:current_site_land_use}
\end{center}
\end{figure}

\noindent \textbf{Reality check evaluation of land use.} Using only the observed share vector $s^{obs}$ and the accessibility-guided allocator, the baseline simulation reproduces the geography of uses with small error (Table~\ref{tab:reality_check}). The share-deviation is $D_{LU} = \sum_{u\in\mathcal{U}}( \hat s_u - s_u^{obs} )^2=5.69\times 10^{-3}$, which corresponds to a mean absolute error (MAE) of 0.02 and RMSE of 0.027. The largest differences are in R (0.043) and I (-0.056), consistent with top-down tiering that defers I/T to coarser tiers and parcel-size guards that shift residuals to R. Minor biases include F (-0.018) and A (0.01); B, E, and G are each within 0.009, and T is within 0.006 of target. Spatially, the fit is convincing (Fig.~\ref{fig:current_site_land_use}). By visual comparison, the baseline closely approaches the ground truth: (1) Commercial (B/F) concentrates at the northeast corner along the primary corridor, matching the observed cluster. (2) Industrial/transportation (I/T) forms a western belt adjacent to the port/rail edge, as in the observed map. (3) Residential (R) dominates the central axis of the study area in both maps. (4) And Parks/Open space (G) fall in locations that closely approximate the observed greens, with similar adjacency to schools and neighborhood streets. Residual mismatches are local and interpretable: small fragments near arterial intersections are smoothed by block geometry, and the $RU \rightarrow R+I$ and $CC \rightarrow B+F$ splits can nudge a few blocks toward R or B when parcel-size thresholds trigger whole-block assignment. Overall, the reality check indicates that the accessibility-guided, priority-queued allocator can recover West Oakland’s prevailing land-use structure with minimal inputs. We therefore proceed to counterfactual policy experiments on this calibrated baseline.

\begin{table}[h]
\caption{Ground Truth Shares vs. Baseline Simulation Shares.}
\label{tab:reality_check}
\begin{tabular}{lcc}
    \toprule
    Use & Target share & Achieved share \\
    \midrule
    F & 0.060 & 0.040 \\
    B & 0.175 & 0.183 \\
    E & 0.045 & 0.056 \\
    G & 0.065 & 0.074 \\
    A & 0.060 & 0.069 \\
    R & 0.330 & 0.376 \\
    I & 0.190 & 0.134 \\
    T & 0.075 & 0.068 \\
    \bottomrule
  \end{tabular}
\footnotetext{\YS{Note: as mentioned in Fig~\ref{fig:current_site_land_use}b, since we have further split the original land use type, the target share of each land use is an approximation.}}
\end{table}

\begin{figure}[H]
\begin{center}
\includegraphics[angle=0,
scale=0.6,
trim={0cm 0cm 0cm 0cm},clip]{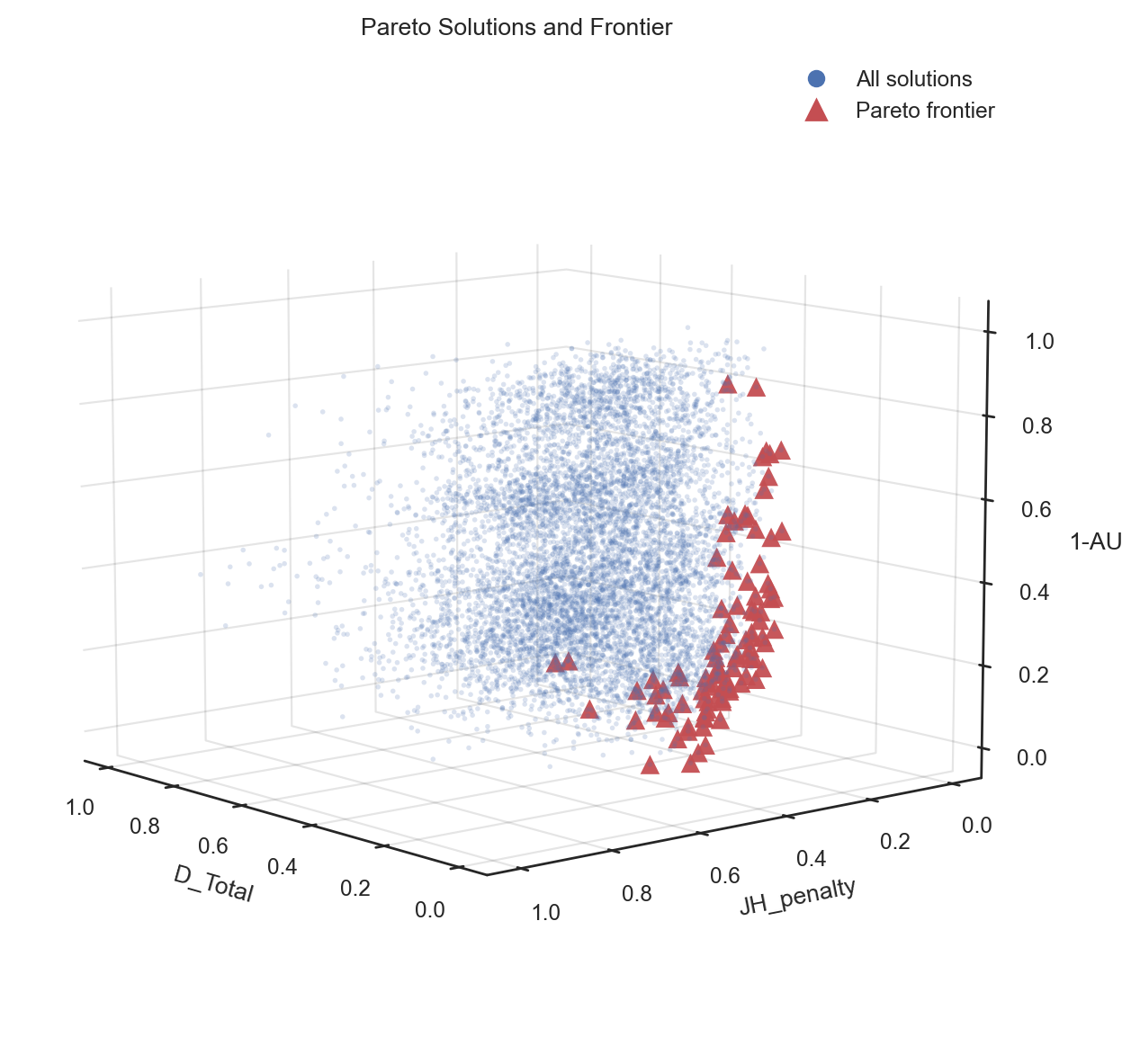}
\caption{3D scatter of the policy design space showing all sampled solutions (circles, faint) and the Pareto frontier (triangles, opaque). Axes represent the three objectives used for non-dominated sorting --- $(1-AU)$, $D_{Total}$, $JH_{penality}$ --- so points nearer the origin are preferred. The frontier traces the best attainable trade-offs among accessibility benefit, aggregate deviation, and jobs–housing balance.} 
\label{fig:pareto_3d}
\end{center}
\end{figure}

\noindent \textbf{Policy search and Pareto analysis.} We evaluate $n_{total}=7,700$ sampled policies and obtain $n_{valid}=7680$ complete records (Fig~\ref{fig:pareto_3d}). Consistent with our setup, we treat accessibility as a benefit and deviations as costs. We compute $1-AU$, and read deviation terms directly from the dataset: block-level area mismatch $D_B$, land-use mismatch $D_{LU}$, construction-share mismatch $D_{CS}$, the jobs–housing penalty $JH_{penalty}$, and the aggregate cost $D_{Total}$. Across all valid policies, ranges are broad: $1-AU \in [0,1]$ (median 0.465), $D_B \in [0,1]$ (median 0.278), $D_{CS} \in [0,1]$ (median 0.420), $JH_{penalty} \in [0,1]$ (median 0.508), and $D_{Total} \in [0,1]$ (median 0.317). Non-dominated sorting on the three-objective vector (($1-AU$), $D_{Total}$, $JH_{penalty}$) yields 79 pareto optimal policies. The frontier markedly improves all metrics relative to the full set: on the frontier, the medians are $1-AU = 0.196$, $D_B=0.016$, $D_{LU}=0.148$, $D_{CS}=0.373$, $D_{Total}=0.142$, $JH_{penalty}=0.321$. This confirms that accessibility gains can be achieved while simultaneously reducing disaggregate mismatches and the aggregate cost term.\\

\begin{figure}[H]
\begin{center}
\includegraphics[angle=0,
scale=0.19,
trim={0cm 0cm 0cm 0cm},clip]{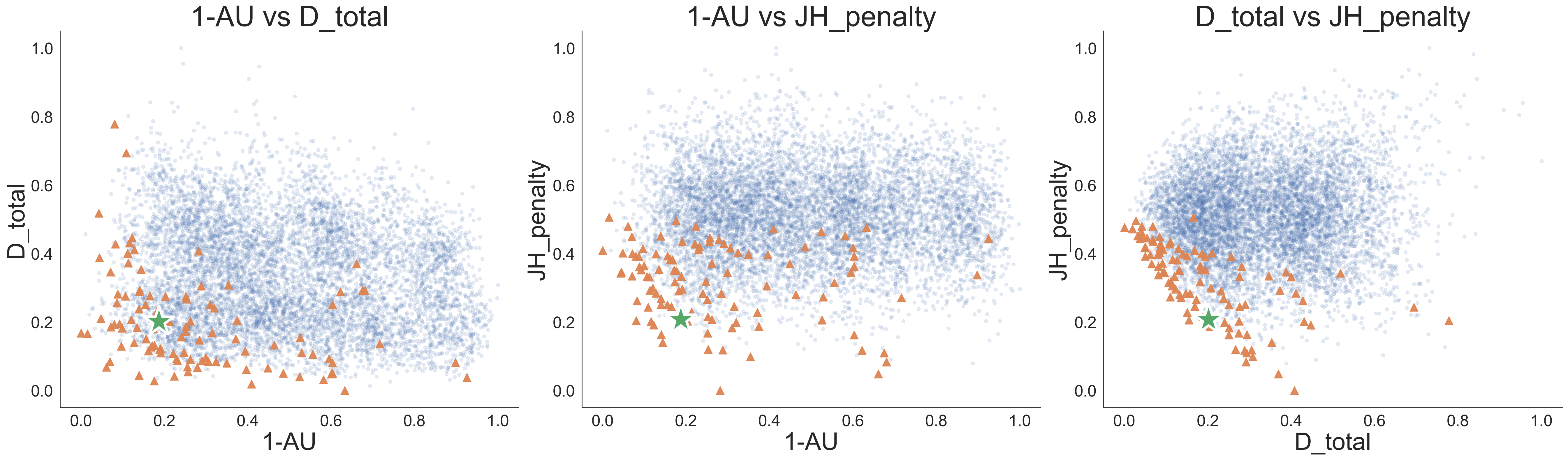}
\caption{Frontier trade-off slices. Two-dimensional trade-off views comparing all sampled policies (faint) to the Pareto frontier (opaque), with the knee solution highlighted (green star). Each panel isolates a pair of objectives to reveal curvature and local trade-offs.} 
\label{fig:fig_tradeoff_slices_2x2}
\end{center}
\end{figure}

\noindent To select a knee, we min–max normalize each objective and choose the policy minimizing Euclidean distance to the utopia point. The knee attains $(1-AU)=0.187$, $D_B=0.003$, $D_{LU}=0.191$, $D_{CS}=0.316$, $D_{Total}=0.201$, and $JH_{penalty}=0.208$. \YS{Its input settings are DistrictRadius=1200, CommunityClusterRadius=900, CommunityRadius=350, (DistanceChoiceWeight0, AngularIntegrationWeight0)=(0.2, 0.8), (DistanceChoiceWeight1, AngularIntegrationWeight1)=(0.2, 0.8), (DistanceChoiceWeight2, AngularIntegrationWeight2)=(0.8, 0.2), with AccTierWeight0=0.5, AccTierWeight1=0.25, AccTierWeight2=0.25.} This policy illustrates the study’s target balance: very high accessibility benefit, very small block-level mismatch, moderate $D_{LU}$, acceptable construction-share deviation, low jobs–housing penalty, and a low composite deviation (Fig~\ref{fig:fig_tradeoff_slices_2x2}).\\

\noindent Across all sampled policies, the aggregate cost $D_{total}$ closely tracks the land-use mismatch $D_{LU}$ (Spearman $\rho=0.86$), while its correlations with the block-area and construction share mismatches are modest ($\rho \approx0.30$ with $D_B$ and $D_{CS}$). Accessibility benefit is only weakly correlated with $D_{Total}$ ($\rho=-0.12$) and jobs–housing imbalance relates mostly to block tightness ($JH_{penalty}$ with $D_B$: $\rho=0.37$) but beneficially to construction composition ($JH_{penalty}$ with $D_{CS}$: $\rho=-0.29$).

\begin{figure}[H]
\begin{center}
\includegraphics[angle=0,
scale=0.42,
trim={0cm 0cm 0cm 0cm},clip]{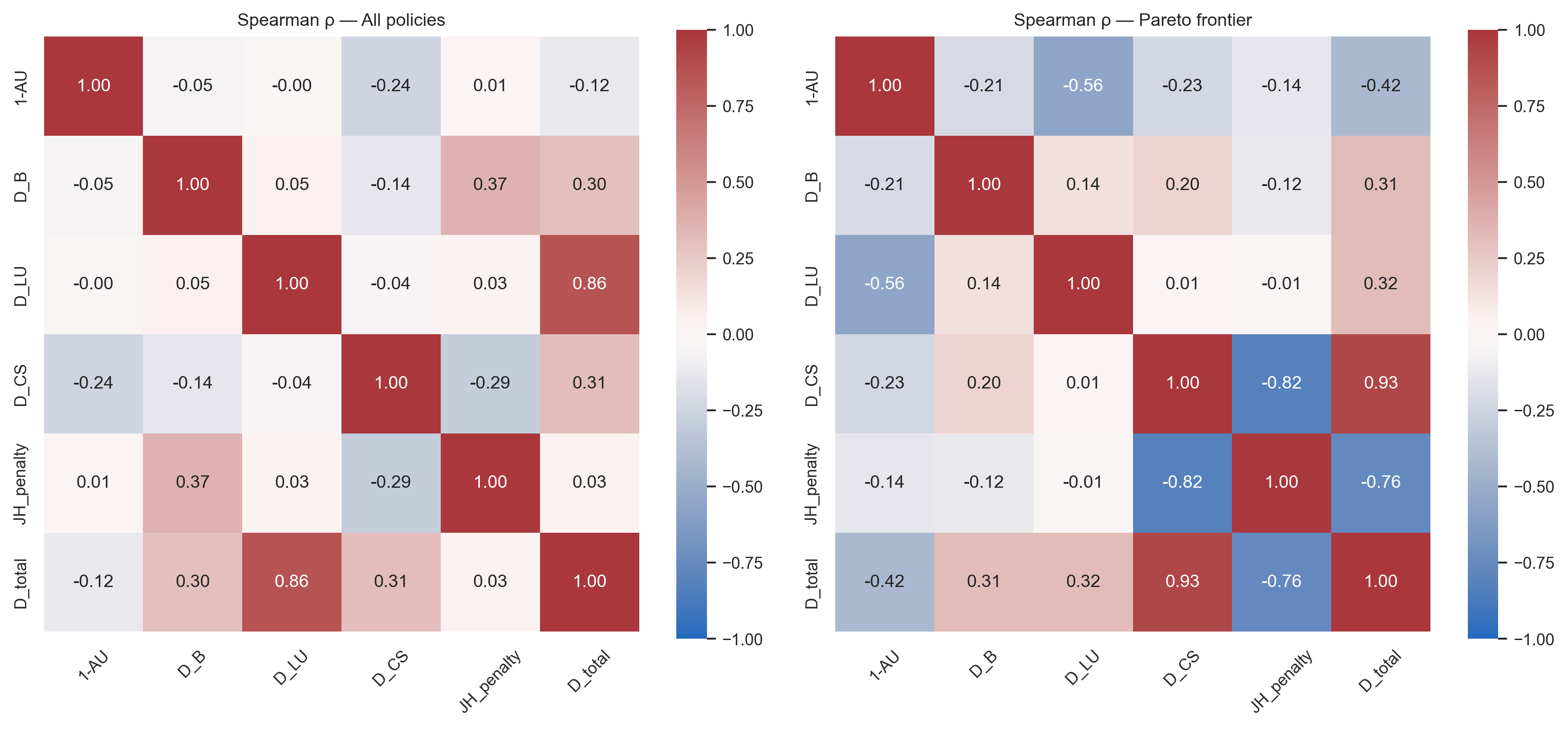}
\caption{Spearman $\rho$ heatmaps. Rank correlation (Spearman) among objectives for (right) all policies and (left) Pareto frontier only. Stronger $|\rho|$ on the frontier indicates tighter structural relationships under optimality.} 
\label{fig:fig_spearman_all_vs_frontier}
\end{center}
\end{figure}

\noindent On the Pareto frontier, these relationships reorganize (Fig~\ref{fig:fig_spearman_all_vs_frontier} right): the dominant driver of $D_{total}$ shifts to the construction-share term, with $D_{CS}-D_{Total}$ rising to $\rho \approx 0.93$, while $D_{LU}-D_{total}$ weakens to $\rho \approx 0.32$. The coupling between accessibility and total cost strengthens (($1-AU$ with $D_{Total}$: $\rho \approx -0.42$), and accessibility aligns most with better land-use matching (($1-AU$) with $D_{LU}$: $\rho \approx -0.56$). Most notably for policy, jobs–housing balance becomes tightly linked to construction composition on the frontier: $JH_{penalty}$ with $D_{CS}$ is strongly negative ($\rho \approx -0.82$), and $JH_{penalty}$ with $D_{Total}$ is also strongly negative ($\rho \approx -0.76$). In short: before optimization, reducing $D_{LU}$ is the surest way to cut $D_{Total}$; after optimizing, the binding margin moves to construction shares—they govern both $D_{Total}$ and $JH_{penalty}$.

\begin{figure}[H]
\begin{center}
\includegraphics[angle=0,
scale=0.35,
trim={0cm 0cm 0cm 0cm},clip]{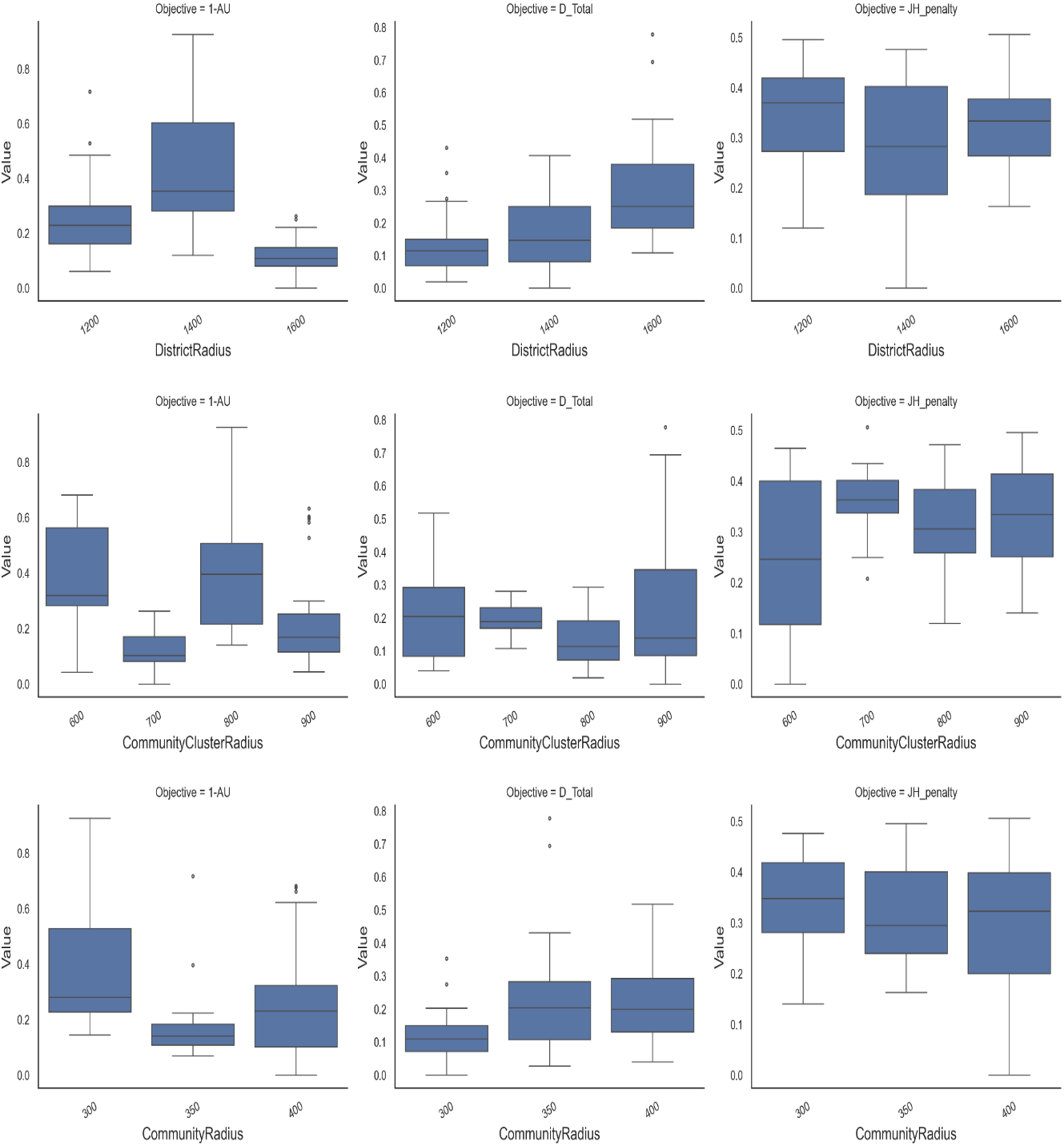}
\caption{Parameter sensitivity on the Pareto frontier. Boxplots of each objective grouped by discrete parameter choices, restricted to Pareto optimal policies. The upper row shows the result grouped by $DistrictRadius \in \{1200,1400,1600\}$, middle row by $CommunityClusterRadius \in \{600,700,800,900\}$, and buttom row by $CommunityRadius \in \{250,300,350,400\}$. Each panel shows how objectives $1-AU$, $D_{total}$, and $JH_{penalty}$ distribute under different parameter choices, revealing which settings consistently improve tradeoffs. Very few frontier policies choose $CommunityRadius=250m$, so we only show the results of other three radii settings.}
\label{fig:Parameter_sensitivity}
\end{center}
\end{figure}

\noindent Fig.~\ref{fig:Parameter_sensitivity} illustrates the sensitivity of the three optimization objectives to variations restricted to optimal policies in district, community, and community cluster radii, with lower values indicating superior performance across all metrics. At the district scale, increasing radii result in elevated median values for $D_{Total}$, suggesting greater structural deviations, though $JH_{penalty}$ declines modestly, reflecting enhanced spatial reconciliation of employment and residential allocations. Larger radii reduce $1-AU$ (improve $AU$) but raise median of $D_{total}$ and $JH_{penalty}$. This trade-off underscores a tension at broader scales: larger district basins facilitate more cohesive jobs-housing distributions by allowing uses to anchor in highly connected corridors, but they amplify deviations from prescribed land-use shares and introduce accessibility inequities. $CommunityClusterRadius$ shows a broad sweet spot around $700-900m$: these settings generally contains relatively smaller $1-AU$, $D_{total}$, and $JH_{penalty}$, while smaller radii are less stable. This pattern shows the advantages of moderately scaled service hierarchies in fostering nested, reproducible placements that avoid unrealistic extremes. It also applies to $CommunityRadius$ tier where scale trade off is revealed: larger radii settings (350-400m) reduce $1-AU$ (improve $AU$) and $JH_{penalty}$ but raise $D_{total}$. Collectively, this suggests that planners should calibrate radii based on policy priorities: for accessibility-driven designs emphasizing equity (low $1-AU$), favor larger district (1400-1600m) and community (350-400m) settings, accepting moderate $D_{Total}$ increases; conversely, for strict adherence to land-use shares (low $D_{Total}$), opt for smaller district radii (1200m) and the 700-900m cluster sweet spot to minimize trade-offs.

\section{Discussion}
The results of this study demonstrate that integrating multi-radius street centralities into a structured workflow can effectively bridge diagnostic accessibility metrics with prescriptive land-use and FAR allocation, yielding plans that enhance urban connectivity while adhering to policy constraints. At the heart of these outcomes lies the mechanism whereby accessibility scores, derived from space-syntax principles, serve as proxies for natural movement patterns and economic viability. For instance, the observed concentration of commercial uses along high-accessibility corridors in the case study district emerges not merely from empirical replication but from the workflow's rule-based prioritization, which favors parcels with superior reach across nested radii. This suggests an underlying causal link: by anchoring land-use decisions to multi-scale centralities, the model amplifies feedback loops between network configuration and programmatic distribution, potentially reducing VMT through localized service basins that align with residents' daily mobility radii. Such dynamics extend beyond surface-level patterns, implying that accessibility-driven allocation mitigates spatial inequities by directing intensity toward nodes that naturally aggregate foot traffic, thereby fostering self-reinforcing urban vitality.\\

\noindent These findings resonate with and expand upon established scholarship in urban morphology and generative planning. Echoing Hillier's "movement economy" paradigm \citep{porta2006network, hillier1996space}, where spatial structure predicates land-use evolution, our pipeline operationalizes this reciprocity by embedding centralities as optimization levers, thus addressing the interpretive gaps in raster-based models like cellular automata \citep{batty1997cellular, batty1997cellularGIS, kelly2021cityengine, white1993cellular}. Unlike procedural grammars in tools such as CityEngine \citep{white2000modeling}, which often impose exogenous intensity envelopes, our approach endogenously derives FAR from accessibility-weighted regressions, challenging the siloed workflows critiqued in prior reviews \citep{ewing2010travel, kockelman1997travel, cervero1996mixed}. The reproduction of industrial belts and corridor-focused commerce in the applied district supports Batty's emphasis on scale-legible metrics for policy evaluation \citep{porta2006network}, yet it extends this by incorporating parcel minima and average-FAR anchors to prevent over-densification extremes, thereby refining the ex-ante testing advocated in land-change simulations \citep{HillierIida2005NetworkPsychUrbanMovement}. Where our results diverge is in the multi-objective Pareto screening, which counters the deterministic biases in single-radius analyses \citep{porta2006network}, revealing trade-offs that prior space-syntax applications have underexplored, such as balancing accessibility gains against deviations in land-share targets.\\

\noindent Theoretically, this work advances three key contributions to urban science. First, it proposes a novel conceptual framework that elevates multi-radius centralities from analytical tools to generative drivers, integrating configuration, program, and intensity into a unified, auditable chain—thus verifying and operationalizing the reciprocal linkages posited in space-syntax theory while revising it to include explicit cluster-aware guarantees for reproducibility. Second, by anchoring FAR estimation to linear models with accessibility weights, the study refines optimization theories in urban design, demonstrating how such mechanisms can reconcile global constraints (e.g., construction totals) with local variances, thereby extending multi-objective paradigms \citep{ewing2010travel} to neighborhood scales. Third, it introduces nested service basins as a mechanism for rule-governed allocation, challenging traditional zoning's rigidity and proposing a hybrid rule-optimization approach that bridges qualitative priorities with quantitative metrics, potentially enriching theories of jobs-housing balance \citep{kockelman1997travel, cervero1996mixed}.\\

\noindent Practically, these insights offer actionable pathways for urban policymakers and designers. For instance, the workflow's transparency enables iterative counterfactual simulations, allowing municipalities to negotiate trade-offs in real-time—such as tilting FAR toward transit-oriented blocks to curb sprawl, as evidenced in the Pareto-efficient plans that minimized VMT deviations. In industry contexts, integration with parametric tools like Grasshopper could streamline district-scale masterplanning, providing developers with reproducible scripts for site-specific adaptations. Policymakers might adopt the multi-objective screening to inform zoning reforms, prioritizing accessibility thresholds in under-connected areas to promote equitable development, while ensuring compliance with environmental targets through adjustable radii.\\

\noindent The study also faces notable limitations that warrant candid scrutiny. The reliance on street-network centralities assumes uniform pedestrian behavior, potentially overlooking barriers like topography or cultural factors that modulate actual accessibility in diverse contexts. Additionally, the model's rule-based land-use placement, while reproducible, may underrepresent stochastic elements in real-world urban evolution, such as market-driven shifts or unforeseen disruptions. \YS{Similarly, the block-level accessibility measure uses only the maximum segment centrality per block. While aligned with a frontage logic, this maxima-based approach can oversimplify connectivity by neglecting blocks with multiple high-centrality frontages. A potential future improvement is a hybrid strategy that retains the intuitive maxima focus but weights it by the proportion of high-centrality segments (e.g., the share of such segments touching a block), thereby capturing a block’s distributed accessibility more comprehensively.} Furthermore, the case study's focus on a relatively small-size district limits generalizability to mega-cities or rural-urban fringes, where multi-modal networks (e.g., rail integration) could alter basin formations. \YS{On interpretability, while we emphasize the procedural transparency of the workflow—where every step from tier weights to priority queues is explicitly defined-we distinguish this from simple predictability. The interaction between multi-scale centrality fields, nested clustering logic, and recursive capacity constraints produces emergent urban forms that are not immediately intuitive. For instance, a minor adjustment in the "District" radius can reconfigure service basins in a way that cascades down to "Community" allocations, shifting commercial corridors by several blocks. Thus, the model functions not as a deterministic calculator, but as a "white-box" complexity engine: it allows planners to audit the logic of allocation (auditability) even if the resulting spatial pattern requires iterative simulation to fully comprehend (emergence).} Finally, computational demands for large-scale applications, including graph processing and optimization sampling, may constrain accessibility for resource-limited planning agencies.\\

\noindent Looking ahead, these constraints illuminate promising avenues for extension. Future inquiries could incorporate agent-based simulations to test the workflow's robustness against dynamic user behaviors, or hybridize it with machine learning to predict land-use transitions under climate scenarios or public health events. Expanding to multi-modal networks would probe its applicability in transit-rich environments, while cross-cultural validations could refine radius calibrations for global contexts. Ultimately, longitudinal studies tracking implemented plans could empirically assess long-term impacts on movement economies, fostering a more resilient urban planning paradigm.

\bibliography{sn-bibliography}

\end{document}